\documentclass[prl,aps,12pt,floatfix]{revtex4-2}

\usepackage{graphicx}
\usepackage{dcolumn}
\usepackage{bm}
\usepackage{siunitx} 
\DeclareSIUnit\torr{Torr}
\DeclareSIUnit{\meV}{meV}
\DeclareSIUnit{\atom}{atom}
\DeclareSIUnit\angstrom{\text{\AA}}
\usepackage[version=4]{mhchem} 

\begin{document}


\title{Probing (sub)nanoscale ferrons in an electron microscope}

\author{Mahir Manna$^{1,2}$}
 
\author{Fei Yang$^3$}
\author{Shayantan Chaudhuri$^{4,5}$}
\author{Sourav Shenoy$^3$}
\author{Guo-Dong Zhao$^3$}
\author{Mohit Tanwani$^6$}
\author{Surya Prakash Reddy$^2$}%
\author{Patrick J. Hays$^2$}
\author{Sriram Sankar$^2$}
\author{Cauê de Souza Coutinho Nogueira$^2$}
\author{Srilekshmi Muraleedharan$^2$}
\author{Xin Xu$^7$}
\author{Sujit Das$^5$}
\author{Katherine Inzani$^4$}
\author{Long-Qing Chen$^3$}
\author{Sandhya Susarla$^2$}
\email{ssusarl3@asu.edu}
\affiliation{%
 1: Department of Physics, Arizona State University, Tempe, 85287, USA
}%
\affiliation{%
 2: Materials Science and Engineering, School for Engineering of Matter, Transport and Energy, Arizona State University, Tempe, AZ 85287, USA
}%
\affiliation{3: Department of Materials Science and Engineering and Materials Research Institute, The Pennsylvania State University, University Park, PA 16802, USA}
\affiliation{%
4: School of Chemistry, University of Nottingham, Nottingham, NG7 2RD, UK
}

\affiliation{%
5: Current address: Department of Chemistry, University of Warwick, Coventry, CV4 7AL, UK
}

\affiliation{%
6: Materials Research Centre, Indian Institute of 5: Science, Bengaluru, 560012, India
}

\affiliation{%
7: The Polytechnic School, Arizona State University, Mesa, Arizona 85212, United States
}

\begin{abstract}
Ferrons are collective excitations of polarization fluctuations \cite{PhysRevB.106.L081105,de1963collective,Choe2026,Wooten2023} that can enable terahertz communications and quantum transduction due to long propagation lengths. Although ferrons have been experimentally demonstrated in van der Waals ferroelectrics \cite{Choe2026} and relaxor ferroelectrics \cite{Wooten2023}, there is no direct (sub)nanoscale experimental evidence of ferrons in three-dimensional ferroelectrics. Here, we detect two types of ferrons, Higgs and pseudo-Goldstone, at the (sub)nanoscale in lead titanate by measuring vibrational signals due to polarization fluctuations. By harnessing momentum transfer in electron energy loss spectroscopy (EELS), we directly distinguish between soft-phonons and ferrons. We observe that the Higgs mode originates from the soft optical phonon parallel to the polar axis, whereas the pseudo-Goldstone mode originates from the soft optical phonon perpendicular to the polarization axis. Together with Landau theory, Raman spectroscopy measurements, and EELS, we observe that Higgs group velocities, in the bulk limit, are eight times greater than those of out-of-plane soft phonons and the pseudo-Goldstone ferrons have group velocities six times greater than in-plane soft phonons due to long-range dipole interactions. We further show that the domain size confinement effects lead to doubling of the respective bulk ferron group velocities, reaching up to approximately \SI{15}{\km\per\second} (almost 15 times higher than the out-of-plane soft phonon modes). Overall, this study opens a pathway to the detection of ferrons in three-dimensional ferroelectrics with domain engineering as a promising avenue for terahertz communication and transduction. 

\end{abstract}

\maketitle

Collective excitations governed by coupled spin, lattice, orbital, and charge degrees of freedom in quantum materials are central to most energy efficient technologies. For example, magnons or collective spin excitations in spintronic devices can route information at low power by moving magnetic excitations rather than electrons~\cite{pirro2021advances}. The question of whether ferroelectric materials that break inversion symmetry support an analogous class of collective excitations, called ferrons, has remained an open challenge~\cite{de1963collective,PhysRevB.106.L081105,PhysRevApplied.20.050501,3y1m-66s1}. 

\begin{figure}[b]
\includegraphics[width=1\columnwidth]{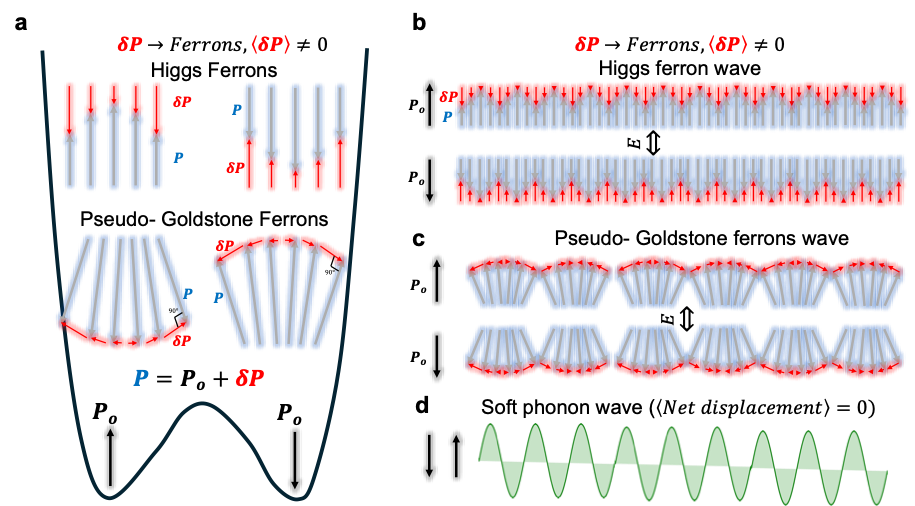}
\caption{\label{FIG 1}\textbf{Concept of ferrons}. \textbf{(a)} Modern Landau double well potential showing that total polarization (blue arrows), P is composed of static polarization (black arrows), $P_o$ and polarization fluctuations (red arrows), $\delta P$. The former is responsible for the structural distortion while the latter results in ferrons. The potential well further shows a half oscillation of Higgs ($\delta P \parallel P$ ) and Goldstone ferrons ($\delta P \perp P$) which changes the total polarization amplitude and direction. Schematic showing the propagation of  \textbf{(b) }Higgs ferron wave, and \textbf{(c)} pseudo-Goldstone wave, whose polarization can be switched from one valley in the Landau potential well to other with electric field ($E$). For cases below the Curie temperature, time averaged polarization fluctuations (\(\langle\delta P\rangle\)) is non-zero. \textbf{(d)} Contrasting behavior in the soft phonon wave where the time averaged atomic displacements from their equilibrium position ( \( \langle\) Net displacement \( \rangle\)) is zero}
\end{figure}

According to Landau theory, ferroelectricity in proper ferroelectrics , such as lead titanate (\ce{PbTiO3}), originates from soft phonons ($\omega_o$) that cause instability in the band structure, leading to tetragonal distortion with two possible polarization directions in the double well potential~\cite{lines2001principles} ($P_o$, black arrows in \textbf{Fig 1a}) . Modern Landau theory expands this framework to include long-range dipole interactions, especially, polarization fluctuations or $\delta P$ (red arrows in \textbf{Fig 1a}) to modify the total polarization, P (blue arrows in \textbf{Fig 1a}) ~\cite{PhysRevB.106.L081105,PhysRevApplied.20.050501}. The new formulation creates ferrons at a unique frequency (\(\omega_{\mathbf{q}}\)) associated with \(\delta P\).  Furthermore, polarization fluctuations  (\(\delta P\)) can be parallel or perpendicular to the total polarization $(P)$, resulting in Higgs and pseudo-Goldstone ferron modes, respectively. Higgs modes reduce the polarization amplitude (See its half oscillation in double well potential in \textbf{Fig 1a} and full wave in \textbf{Fig 1b}) , whereas pseudo-Goldstone modes introduce a net rotation in the polarization vector (See its half oscillation in double well potential in \textbf{Fig 1a} and full wave in \textbf{Fig 1c})). It should be noted that Goldstone modes are gapless excitations that require the existence of a Mexican hat type potential~\cite{RevModPhys.95.011001}. This potential is not possible in proper ferroelectric systems, such as PbTiO$_3$, due to restrictions in dipole interactions created by the fixed positions of the atoms. Thus, to distinguish the present Goldstone ferrons from conventional Goldstone modes, we use the terminology of `pseudo-Goldstone ferrons'. Below the ferroelectric transition, time averaged polarization fluctuations is non-zero , which is the key distinction from soft phonons (See \textbf{Fig 1b-d}). Finally, the direction of the polarization fluctuations can be switched with an electric field (E), which makes it exciting for terahertz communications and quantum transduction. 

Both ferrons and soft phonons originate from similar atomic displacements; therefore, at the $\Gamma$ point, $\omega_o$ and \(\omega_{\mathbf{q}}\) frequencies will match. However, at different momentum transfers, both frequencies will separate due to different group velocities. Ferrons are predicted to propagate faster than soft and acoustic phonons due to long-range dipole-dipole interactions~\cite{3y1m-66s1}. The long-range dipole interactions also changes the ferron energy positions at different domains, analogous to hybrid light-matter excitations \cite{basov2025polaritonic}.  Most phonon-based measurements on three-dimensional ferroelectrics have been performed using Raman spectroscopy~\cite{Burns1973} that has limited spatial resolution and restricts information only to the $\Gamma$-point. This is one of the key reasons why ferrons have not been directly observed experimentally in three-dimensional ferroic materials. 

As a first step to recognize ferrons, the polarization textures in thin films must be re-examined. A \ce{PbTiO3} thin film epitaxially grown on a dysprosium scandate (\ce{DyScO3}) substrate via reflection high-energy electron diffraction (RHEED)-assisted pulsed laser deposition (\textbf{see Fig SI 1}) imparts 1.30\% tensile strain. This creates two domain types: c- and a-, where the polarization and bond elongation are along $\langle001\rangle_\text{pc}$ and $\langle100\rangle_\text{pc}$, respectively, and the domain wall is oriented at \SI{45}{\degree} with respect to the substrate-thin film interface~\cite{nesterov2013thickness}. Our experiments confirm this structure both at the mesoscale (\textbf{Fig 2a}) and at high resolution (\textbf{Fig. 2b}) without the presence of oxygen vacancies (\textbf{Fig SI 2}). When moving from c- to a- domain, the polar axis rotates by \SI{90}{\degree}. However, the tetragonality (polar axis to non-polar axis lattice parameter ratio) is similar for both c- and a-domains (indicated by the colorbar in \textbf{Fig 2b}) . 

\begin{figure}[t]
\includegraphics[width=1\columnwidth]{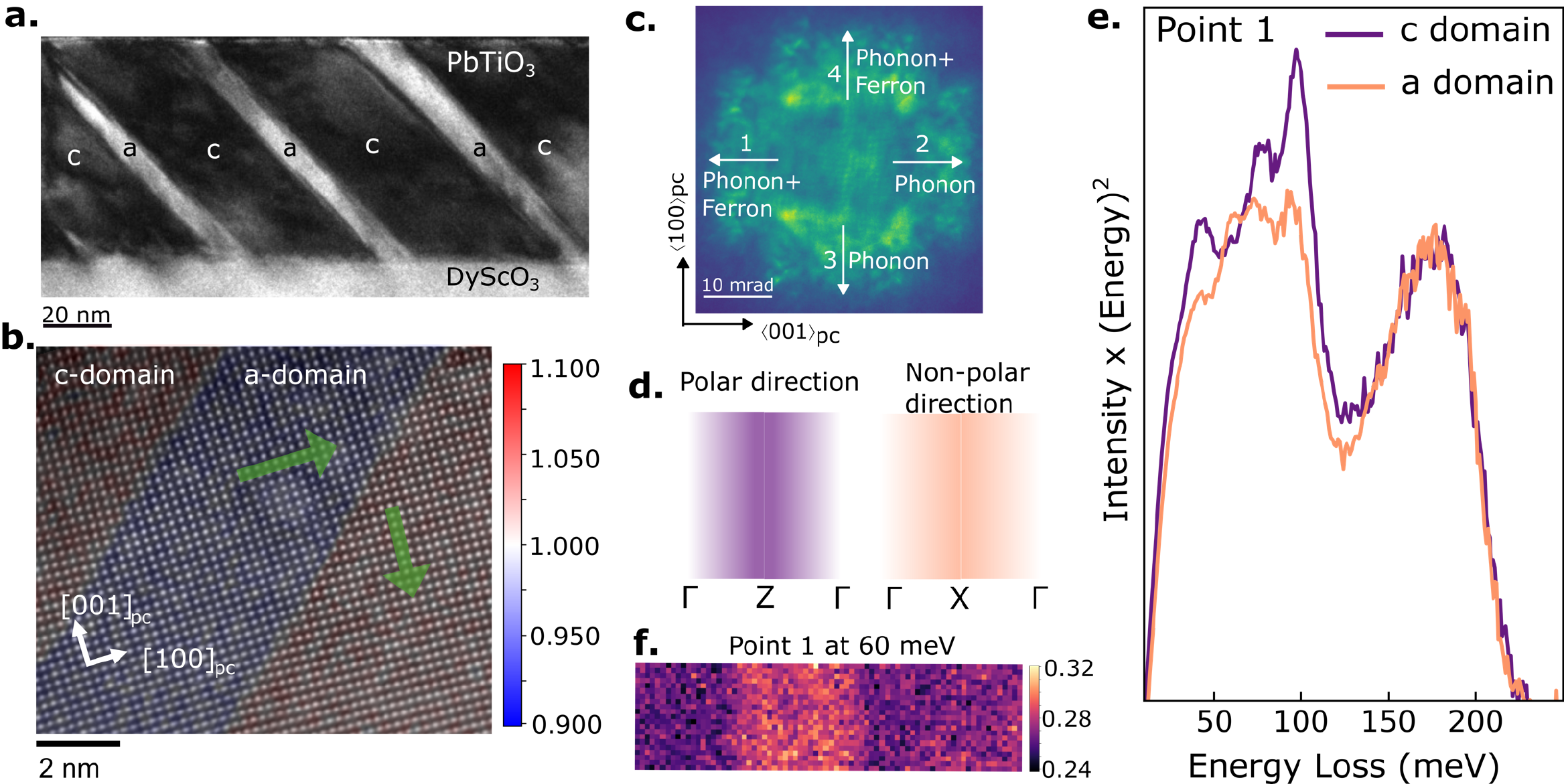}
\caption{\label{FIG 2} \textbf{Vibrational mapping in Lead Titanate.} \textbf{(a) }Weak-beam dark-field transmission electron microscopy (TEM) image of a \ce{PbTiO3} thin film grown on a \ce{DyScO3} substrate, showing c-domains (extended regions) and a-domains (narrow bright stripes).\textbf{ (b)} Atomic-resolution high-angle annular dark-field scanning TEM image; the green arrow indicates the orthogonal polarization directions in the c- and a-domains. \textbf{(c)} Experimental ronchigram of \ce{PbTiO3}, where the white arrows indicate the direction in which the bright field disk is shifted,. (1,2) are conjugate pairs along $\langle001\rangle_\text{pc}$ and (3,4) are along $\langle100\rangle_\text{pc}$ where is pc denotes pseudo cubic. \textbf{(d)} Effective momentum transfer due to convergent angle, collection angle settings and dynamical electron scattering in \ce{PbTiO3}, where the increase in the color gradient denotes in the increased dominance of the particular momentum transfer along polar (magenta) and non-polar (orange) axis.  \textbf{(e) }IE$^2$ representations of off-axis vibrational EEL spectra acquired from the c- and a-domains along the $\langle001\rangle_\text{pc}$ direction, showing clear spectral differences, particularly at the \SI{60}{\milli\electronvolt} and \SI{97}{\milli\electronvolt} peak positions. \textbf{(f) }Spatial mapping based on the \SI{60}{\milli\electronvolt} vibrational peak in the EEL spectra acquired along $\langle001\rangle_\text{pc}$ direction, clearly resolving and distinguishing the c-(magenta) and a-domain (orange) regions.}
\end{figure}

Since ferrons are vibrational-type excitations, they can be measured using scanning transmission electron microscopy electron energy loss spectroscopy (STEM-EELS), which uses a convergent electron beam to scan the material and collect inelastic scattering signals at every position. The key to ferron detection is to identify vibrations caused by ferroelectric polarization fluctuations. The traditional way to characterize polarization is to compare the intensities of Friedel pair disks~\cite{bird1989phase}. This should probe $P$, which has both $P_o$ and $\delta P$ components. The $P_o$ (static polarization) will cause a distortion in the structure, as frequently observed by four-dimensional (4D) STEM~\cite{bird1989phase}. Since this is a static feature, its long-range order cannot result in any new fluctuations unless the polar phonons are directly coupled to the electron beam, resulting in phonon-polaritons~\cite{bib17} (\textbf{see Supplementary Note 2.1,2.2}). The polarization fluctuation, $\delta P$  cannot be visualized via 4D STEM as it requires time resolution, but its time-averaged frequency can be captured via inelastic scattering, thus directly allowing ferrons to be probed (\textbf{Details in supplemental Note 2.1,2.2})

With this as motivation, we perform convergent differential momentum $(q)$-averaged  STEM-EELS in \ce{PbTiO3}. The principle behind this technique is to shift the convergent beam in both polar- and non-polar conjugate directions in the diffraction space (\textbf{Fig 2c}) where the differences in the conjugate pairs can give direct access to ferrons (\textbf{Details in Supplemental Note 2.1}) . It should be noted that differential EELS-type geometry has previously been attempted~\cite{Gadre2022a,hoglund2025topologydrivenvibrationschiralpolar,gadre2022nanoscale}. However, this is the first time that it has been used with a higher convergence angle to recognize ferron interactions by controlling momentum transfers that are complementary to recent observations of thermal ellipsoids~\cite{YanNature25}. The rotation calibration (\textbf{Fig SI3}) permits both polar and non-polar directions to be identified in \ce{PbTiO3} (\textbf{Fig 2c}). Perovskites undergo dynamical electron diffraction, due to which the interpretation of momentum transfers is not the same as in conventional systems, such as silicon~\cite{pfeifer2026phononselectioninterferencemomentumresolved}. For the current setup, where the convergence angle is \SI{19}{\milli\radian}, the collection angle is \SI{20}{\milli\radian} and the beam shift is \SI{40}{\milli\radian}, the effective momentum transfer enables the detection of vibrations, mainly from a narrow band near the X-point in the non-polar direction and near the Z-point in the polar direction (\textbf{Fig 2d, Supplemental Note 2.3,3, Fig SI4-6}). Next, two-dimensional vibrational mapping was performed (See details in \textbf{Methods}) along each conjugate pair direction. An example of  convergent differential $q$-averaged  STEM-EELS mapping along the $\langle100\rangle_\text{pc}$ direction is shown in \textbf{Fig 2e,f}, where the domains in \ce{PbTiO3} have been identified due to differences in the phonon density of states. Next, the ferrons signals were extracted from the spatially-averaged raw vibrational signals using the following steps: 1) pre-post edge background subtraction, 2) bulk phonon-polariton removal, and 3) linear combination fitting of the conjugate pairs. The edges of the c- and a- domains, which is $\approx$ 3 nm region around the domain wall was intentionally not considered during spatial averaging to avoid complications related to phonon-polaritons at the domain walls.  Each step in the entire process accumulates uncertainties, which are cumulatively benchmarked against strontium titanate (\ce{SrTiO3}), a centrosymmetric material, where no ferrons are expected (Details are in \textbf{Supplemental Note 4}).  

\textbf{Fig 3a} shows the residuals from conjugate pairs in \ce{SrTiO3}, shown as white background, and in \ce{PbTiO3}, along both polar- (magenta) and non-polar (orange) axes in the c- and a-domains. Calibration of the data processing method using SrTiO$_3$ indicates that the first peak in the residual PbTiO$_3$ is highly susceptible to data processing artifacts. Hence, we do not interpret this signal here. Ignoring these uncertainties, the c-domain in \ce{PbTiO3} shows the presence of a broad peak at approximately \SI{57}{\milli\electronvolt} and \SI{46}{\milli\electronvolt} in the polar and non-polar directions, respectively. These observations were also repeatable when point spectra averaged from 20 randomized points in the c-domain were acquired under similar conditions and underwent the same data processing routine (\textbf{Supplemental Note 5}). To understand the origin of these peaks, the selection rules of Higgs and pseudo-Goldstone ferrons were studied (\textbf{See Supplemental Note 2.3}). Higgs ferrons are parallel to the polarization and therefore are only permitted along the polar axis (see \textbf{Fig 1a} schematic). Pseudo-Goldstone modes, however, are perpendicular to the polarization and therefore are only permitted along the non-polar axis  (see \textbf{Fig 1a} schematic). These experimental observations indicate that the peaks along the polar axis marked with * are Higgs modes and * peaks along the non-polar axis are pseudo-Goldstone modes. We further validate these observations with Landau theory (\textbf{Supplemental Note 7}).

\begin{figure}[htbp]
\includegraphics[width=0.7\columnwidth]{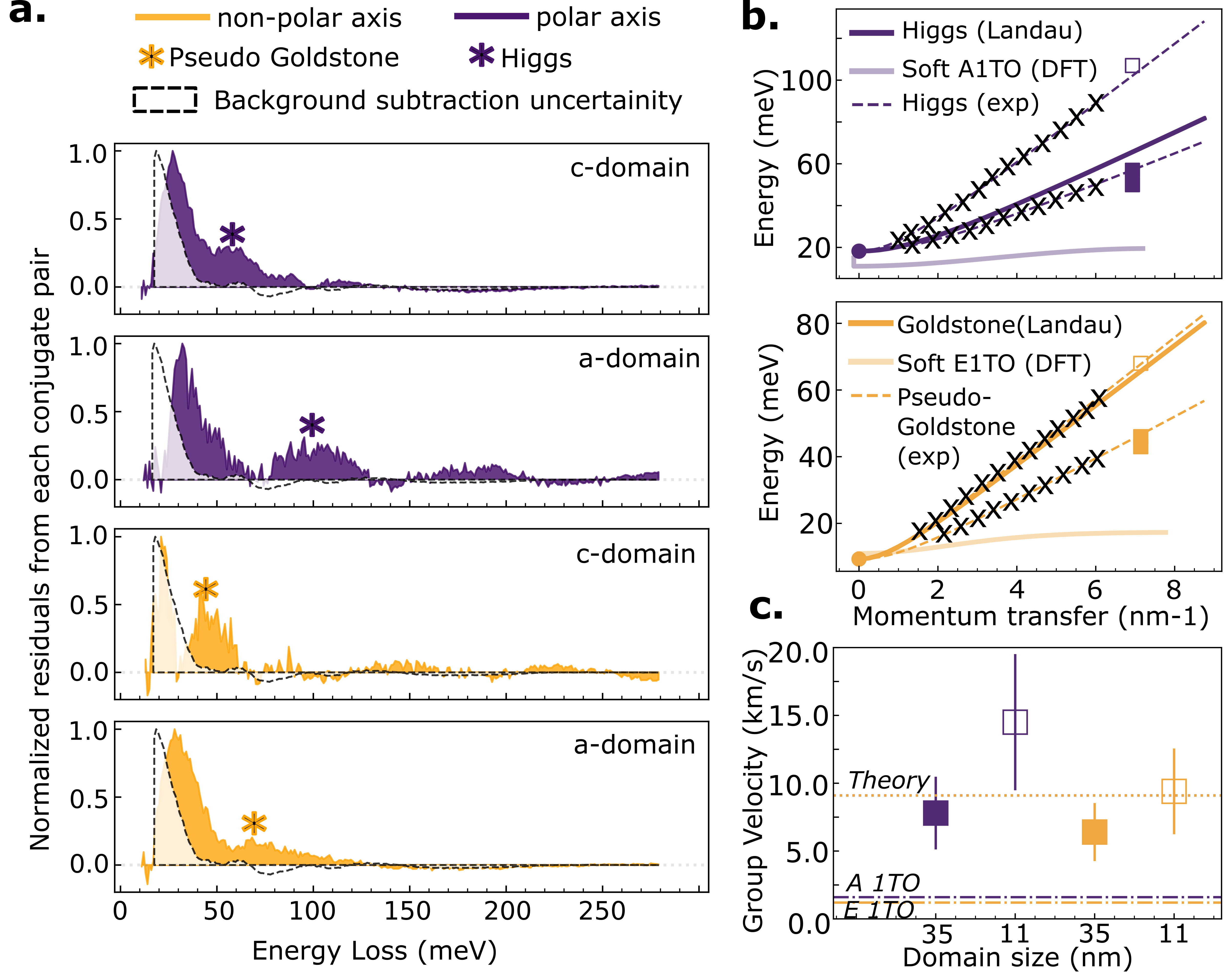}
\caption{\label{FIG 3} \textbf{Discovering Higgs and pseudo-Goldstone ferrons, and their dispersion.} (a) Residuals from the conjugate pairs in c- and a-domains along the polar (magenta) and non-polar (orange) axes in \ce{PbTiO3}. The white spectra overlaid on top represents the background subtraction uncertainty extracted from average of residual of two \ce{SrTiO3} conjugate pairs. Ferrons corresponding to polarization fluctuations are indicated as *. (b) Higgs (magenta) and pseudo-Goldstone (orange) ferron dispersion relations compared with soft phonon dispersions extracted from the $\Gamma-X$ (translucent orange) and $\Gamma-Z$ points (translucent magenta), respectively. The solid line is the ferron dispersion relation from Landau theory and the dashed line is the experimental curve fit using Raman spectroscopy (circles) and EELS (solid squares c-domain and hollow squares a domain). The "X" on the curve fit denotes the points where the ferrons signals are inaccessible due to the limitations in spatial and spectral resolution in both Raman and EELS experiments. (c) Experimental Higgs (magenta) and pseudo-Goldstone (orange) group velocities extracted from experiments. Errors bars originate from the momentum transfer bands in the current experiments. The colored dotted line represents the  Higgs and pseudo-Goldstone group velocities from Landau theory. The dash dotted line is the soft optical phonon group velocities extracted from density functional theory: A1 (1TO) in magenta and E (1TO) in orange.}
\end{figure}

Landau theory predicts the bulk ferron properties of bulk PbTiO$_3$ ignoring microstructural details such ferroelectric and ferroelastic boundary conditions that affects c- and a-domains in thin films. It is therefore expected that the ferron energy positions and dispersion predictions from Landau theory may be close to experiments, but may not be an exact match. It starts with an effective Hamiltonian with static polarization and polarization fluctuations, especially taking into account quadratic and sixth order anharmonic interactions, all of which are critical to resolve the ferron spectrum and dispersion. The anharmonic coefficients, the collective-mode gaps, and the polarization fluctuation amplitudes are mutually coupled, forming a closed set of self-consistent equations. Starting from the zero-temperature parameters of quadratic and sixth order anharmonic interactions, effective inertia (mass) of the polarization field ($m_p$), and stiffness coefficient of the polarization field ($g$) ,  these equations iteratively at each temperature (\textbf{Supplemental Note 7, Fig. SI 16}). 

To compare with the current experiments, room temperature parameters are used. The ferron gap and soft phonons energies match with each other. We validate the temperature dependent ferron gap curves by comparing it with temperature dependent soft phonons frequencies along non-polar and polar directions using Raman spectroscopy (\textbf{Supplemental Note 6, Fig SI 15}). The ferron Higgs and pseudo-Goldstone ferron gaps match soft transverse optical phonons along the polar and non-polar axes, respectively. (\textbf{Supplemental Note 7, Fig. SI 16}). Furthermore, even the ferroelectric transition temperature calculated by these higher order anharmonic coefficient is validated by phase field simulations (\textbf{Supplemental Note 8}) which matches to a resonable extent. Next, the group velocity at room temperature was calculated by ratio of stiffness coefficient of the polarization field ($g$) to effective inertia (mass) of the polarization field ($m_p$), which was found to be 9.1 km/s. The ferron gap and the group velocity was taken together to calculate the dispersion relation according to equation 4 in the methods section (\textbf{Fig 3b}). 

Experimentally extracting the ferron dispersion is challenging because microstructural sensitivity requires a STEM probe size below one unit cell dimension, and the expected ferron gap (approximately 9-18 meV) lies near approximately 15 meV spectral resolution of off-axis EELS. Calibration on SrTiO$_3$ indicates that EELS can reliably access ferron energies only above approximately 40 meV, corresponding to minimum effective momentum transfer of approximately 6 cm$^{-1}$ in the dispersion relation (X-marked regions in \textbf{Fig. 3b}). So, four-dimensional STEM-EELS cannot be used here \cite{WU2023113818} as it compromises spatial resolution and also cannot resolve the ferron gap due to limited spectral resolution in off-axis settings. We therefore combine Raman and EEL spectroscopy to access the full dispersion. The $\Gamma$-point frequency was extracted from Raman spectroscopy, while the non-$\Gamma$-point frequency was extracted from the spatially averaged conjugate pair residual spectra in EELS.  Since EELS uses a convergent beam, it is sensitive to a finite range of momentum transfer rather than a single point (\textbf{See Supplemental Note 2.3}). For the current conditions in \ce{PbTiO3}, the experimental momentum transfer vectors along the polar and non-polar axes were estimated to be \SI[separate-uncertainty=true]{6.9(2.16)}{\per\nm} and \SI[separate-uncertainty=true]{7.1(2.2)}{\per\nm}, respectively. The experimental (dotted lines) and theory dispersion (solid lines) curves for c-domain closely resemble each other. Further insights were gained by calculating the ferron group velocity. 

Experimental group velocities were extracted using the momentum transfer band and compared with Landau theory calculations. The experimental Higgs group velocity in the c-domain was approximately \SI[separate-uncertainty=true]{7.8(2.2)}{\km\per\second}; the error arises from the band of momentum transfers in EELS. The Higgs mode group velocity is almost eight times higher than that of A1 (1TO) soft phonons (approximately \SI{1.7}{\km\per\second}, as extracted from density of states, \textbf{Fig SI6}). The pseudo-Goldstone mode group velocity in the c-domain, approximately \SI[separate-uncertainty=true]{6.4(2.2)}{\km\per\second}, is almost six times higher than that of the E (1TO) soft phonons (approximately \SI{1.2}{\km\per\second}, as extracted from density of states, \textbf{Fig SI6}). Both experimental ferron group velocities are slightly lower than the expected theoretical value of \SI{9.1}{\km\per\second}. The minor discrepancy may arise from the current limitations in Landau theory which doesnot include the complete microscopic interactions and also from the spread in the momentum transfers in EELS. The group velocities higher than the optical phonon branches but lower than photons suggest hybrid light-matter coupling characteristics. 

An important feature of hybrid light-matter excitations is the sensitivity to confinement effects. Ferron frequencies will be sensitive to different polarization domain sizes, analogous to how polaritons behave in confined interfaces~\cite{basov2025polaritonic}. Bulk (an)harmonic phonons, however, will show little dependency on thickness. We observe domain-size-dependent hardening (\textbf{Fig 3a}) consistent with confinement of the collective excitation. Ferrons, marked using *, display hardening as the polarization textures become smaller. As seen in \textbf{Fig. 3a}, the peak shifts from approximately \SI{57}{\milli\electronvolt} to \SI{106}{\milli\electronvolt} from the c- to a- domain along the polar axis. Since both c- and a-domains possess the same tetragonality, the $\Gamma$-point soft phonons (which are the same frequency as ferrons at $q=0$) will not change. Thus, the same $\Gamma$-point frequencies from the c-domain in Raman spectroscopy can be used for the a-domain. The non-$\Gamma$-point ferron frequencies in the a- domain are extracted from STEM-EELS. The group velocity of the Higgs mode in the a-domain was approximately \SI[separate-uncertainty=true]{15(5)}{\km\per\second}, which is almost double that of the Higgs group velocity in the c-domain and almost 15 times that of the A1 (1TO) phonon (\textbf{Fig 3c}). The pseudo-Goldstone mode also doubles its c-domain group velocity, reaching up to \SI[separate-uncertainty=true]{9.4(3.2)}{\km\per\second}, almost 9 times higher than that of the E (1TO) phonon (\textbf{Fig 3c}). 

\section*{Discussion and Conclusions}

This work establishes a general micro-structural framework for isolating Higgs and pseudo-Goldstone ferrons in three-dimensional ferroelectrics, distinguishing them from anharmonic soft-phonons via momentum-averaging at non-Gamma points, probing across different polarization textures, and experimentally extracting the unique dispersion of ferrons. We directly probe that Higgs ferrons are in the same direction as ferroelectric polarization and that pseudo-Goldstone modes are perpendicular to the ferroelectric polarization direction, which can be cleverly used to electrically control the ferron via device electrodes that can specifically control out-of-plane polarization. The ferron detection approach extends directly to the likes of lithium niobate, barium titanate, and other relaxor ferroelectrics, where such ferrons have already been predicted~\cite{PhysRevB.106.L081105}. Both Higgs and pseudo-Goldstone exhibit hybrid light-matter character, with group velocities exceeding those of optical phonons in \ce{PbTiO3}, giving rise to quantum confinement effects. Although only two thicknesses are examined here, this points to cavity-like ferron behavior achievable through domain engineering of oxide heterostructures~\cite{schlom2007strain}. Notably, a 31\% reduction in polarization texture size nearly doubles the bulk group velocity, indicating long propagation lengths both along and perpendicular to the polarization direction. This size-dependent tunability offers a design principle for terahertz-range interconnects: electrically written and erased nanoscale domains could enable long-distance terahertz signal propagation via ferron waveguides, analogous to polaritonic waveguides in the visible range \cite{Stoyanov2002}

There are also several open questions from this work. For example, it is unknown why the ferron group velocity increases with the decrease in domain size, but it is in agreement with the hypersonic ferron velocities recently observed in atomically thin van der Waals ferroelectrics \cite{Choe2026}. More first-principle-based modeling and experimental domain size dependent ferron measurements are needed to conclude the physical mechanism behind this behavior. Next, several phonon modes in \ce{PbTiO3} are chiral~\cite{Yang2026}. How chirality affects subsequent ferron behavior is currently unknown. This would require changing sample preparation and the electron beam shifting in off-axis STEM-EELS. This is currently possible and has already been proven in probing orbital and spin angular momentum resolved electronic states in magnets~\cite{Ali2025,Song2026}, and could be extended to phonons and ferrons. There is also the possibility of multi-ferrons existing, where the ferrons themselves also carry magnetic moments~\cite{2sj3-33ky}. The detection of ferron chirality will also provide experimental proof for multi-ferrons. Overall, this work marks a new era for nanoscale probing of ferrons, complementary to recently probed nanoscale magnons~\cite{Kepaptsoglou2025}, that could be extended to other ferroelectric systems~\cite{mr9v-khtj, Choe2026, xu2020strain,wang2003epitaxial,yadav2016observation, das2019observation,sanchez20242d,hadjimichael2021metal} and performed under different external stimuli such as at liquid helium temperatures~\cite{zhang2026nanoscalepolarlandscapesquantum} or under \textit{operando} electric field~\cite{Liu2025}. 
\clearpage

\begin{acknowledgments}
\textbf{Acknowledgements} S.Susarla, S. Sankar, P. J. H., S.M. and M.M. acknowledge the use of Arizona State University startup funds. M.M was partially supported by the Scialog Program by the Research Cooperation for Scientific Advancement. S.Susarla has been partially supported by the Army Early Career Program under award number: W911NF-26-1-A047. S.C. acknowledges support from Wellcome Leap as part of the Quantum for Bio Program. The efforts of F.Y., S.Shenoy., G.D.Z, and L.Q.C. are supported as part of the Computational Materials Sciences Program funded by the US Department of Energy, Office of Science, Basic Energy Sciences, under Award Number DE-SC0020145. K.I. acknowledges support from the EPSRC Fellowship program [EP/W028131/1]. We acknowledge the use of facilities within the Eyring Materials Center, Arizona State University, supported in part by NNCI-ECCS-1542160. Computing resources were provided by the University of Nottingham, the EPSRC-funded HPC Midlands+ consortium [EP/T022108/1] for access to Sulis, the EPSRC-funded High-End Computing Materials Chemistry Consortium [EP/X035859/1] for access to the ARCHER2 UK National Supercomputing Service~\cite{BeckettZenodo24_ARCHER2} (\url{https://www.archer2.ac.uk}), the National Academic Infrastructure for Supercomputing in Sweden, and the Swedish National Infrastructure for Computing at the National Supercomputer Centre in Sweden which was partially funded by the Swedish Research Council [Grant Agreement Nos. 2022-06725 and 2018-05973]. M.T. and S.D. acknowledge support from the Fund for Improvement of S\&T Infrastructure (FIST-DST) grant [SR/FST/ET-II/2023/1233 (C)], the Infosys Young Investigator Award and an Indian Institute of Science start-up grant. We acknowledge Shize Yang, Piyush Haluai, and Manuel Roldan Gutierrez (Eyring Materials Center, Arizona State University) for helping with microscope training and sample loading. We also thank Peter Crozier and Yifan Wang (School for Engineering of Matter, Transport and Energy, Arizona State University) for their scientific inputs. We thank Eric Hoglund (Oak Ridge National Laboratory) and Amber Quillin (Bruker AXS) for sharing their tips and tricks for collecting off-axis STEM-EELS datasets. Finally, we thank Nicola Spaldin (ETH Zurich), and Jan Rusz (Uppsala University) for providing valuable the technical feedback. 

\noindent \textbf{Author Contribution:}  S.Susarla conceived the idea. M.M. performed polarization dependent Raman measurements, off-axis STEM-EELS and convergent differential momentum $(q)$-averaged  STEM-EELS experiments and the majority of data interpretation. S.Susarla performed HAADF-STEM measurements and STEM-EELS data analysis. S.C. and K.I. performed DFT calculations. M.T. and S.D. grew the \ce{PbTiO3} thin films. S.P.R. assisted in the optimization of the background subtraction.  P.J.H. assisted in the interpretation of the datasets. S.M. performed weak-beam dark-field measurements. F. Y calculated the ferron spectrum using Landau theory. G.D and S. Shenoy performed the phase field simulations. S.Sankar assisted with preliminary temperature dependent Raman measurements. C.N performed the temperature dependent Raman measurements. P.R helped with the dielectric formulation theory in EELS. K.I., L.Q.C.,  X.X., and S.Susarla. supervised the project. X.X, K.I., L.Q.C., S.Susarla. and S.D. obtained funding for the project. M.M, F.Y, S.C and S.S wrote the initial version of the manuscript. All authors contributed to the revision of the manuscript.

\end{acknowledgments}

\appendix

\section{Appendixes}

\noindent\textbf{Density Functional Theory Calculations}
Kohn-Sham density functional theory~\cite{HohenbergPR64, KohnPR65} calculations were performed using version 6.3.0 of the VASP~\cite{kresse1996efficient, kresse1999ultrasoft, kresse1996efficiency} software, with the Pb\_d ($5d^{10}6s^26p^2$), Sr\_sv ($4s^24p^64d^{0.001}5s^{1.999}$), Ti\_sv ($3s^23p^63d^34s^1$), and O ($2s^22p^4$) projector-augmented wave pseudopotentials. The PBEsol~\cite{PBEsol} generalized gradient approximation and a converged plane-wave energy cutoff of \SI{750}{\electronvolt} was used in all calculations. Furthermore, a converged reciprocal space grid ($k$-grid) of size $6\times6\times6$ was used for both \ce{PbTiO3} and \ce{SrTiO3} primitive unit cells. Convergence tests for the plane-wave energy cutoff and the $k$-point mesh were performed until the total energies from static calculations were converged to within \SI{1}{\meV\per\atom} (\textbf{Supplemental Figures SI-7 and SI-8}). Both unit cells were subsequently optimized using the converged settings until the residual forces on all ions were less than \SI{0.01}{\electronvolt\per\angstrom}. 

The Phonopy~\cite{TogoJPCM23_Phono3py, TogoJPSJ_Phono3py} package was used to perform harmonic phonon calculations on $4\times4\times4$ supercells of the PbTiO$_3$ and SrTiO$_3$ unit cells. Atomic forces were calculated from finite displacements of the supercells and used to calculate second-order force constants. A non-analytical correction term was applied to correct phonon band structures, since atomic displacements can induce polarization in non-metallic crystals and the resulting macroscopic field can lead to LO-TO splitting near the $\Gamma$-point of the first Brillouin zone. A $71\times9\times9$ $q$-mesh (where $q$ denotes the phonon wavevector in reciprocal space) was employed to compute phonon density of states spectra and output frequencies. The non-analytical term correction was applied along the $(0,0,0.1)$ direction (in reciprocal lattice units), which corresponds to a small displacement from the $\Gamma$-point along the [001] direction. Convergence tests were carried out on both unit cells to determine the dielectric tensor and appropriate $k$-grid size for the non-analytical correction (\textbf{Supplemental Figures SI-9 and SI-10}). Based on Frobenius norm comparisons between successive meshes, $k$-grids of size $9\times9\times8$ for PbTiO$_3$ and $9\times9\times9$ for \ce{SrTiO3} were selected. The Sumo~\cite{GanoseJOSS18_Sumo} toolkit was used to plot and analyze phonon spectra. 
\\

\noindent\textbf{Self-Consistent Field Calculations of Collective Modes} 
The ferroelectric system can be described by an effective lattice-dynamical Hamiltonian associated with the transverse soft phonon mode~\cite{cochran1981soft,cowley1996phase,cowley1965theory}. Upon cooling, the transverse soft phonon frequency continuously decreases and eventually becomes imaginary at the center of the Brillouin zone~\cite{PhysRevB.111.094113,verdi2023quantum,he2020anharmonic}. This instability signals the onset of a macroscopic condensation of the soft-phonon field at $q=0$, analogous to the Bose-Einstein condensation of bosonic particles. Within the mean-field approximation, the condensation leads to a nonzero expectation value of the zone-center soft-phonon field, $
\xi=\langle \phi_{\rm sp}(\mathbf{q}=0)\rangle$~\cite{lee1974conductivity}, 
which corresponds to a collective lattice distortion and generates a spontaneous polarization field ${\hat{\bf P}}=({z^*{\bf e}_{\bf u}}/{\Omega_{\rm cell}}){\hat \xi}$. Here ${\bf e_u}$ is the ionic unit vibration vector, $z^*$ is the effective charge, and $\Omega_{\rm cell}$ denotes the unit cell volume. Then, for the single crystal of \ce{PbTiO3}, the symmetry-allowed ground-state Hamiltonian of the long-range polarization field can be written as~\cite{lines2001principles}:
\begin{align}
&\mathcal{H}=\int \Big[\frac{m_p}{2}(\partial_t{\hat {\bf P}})^2+\frac{g}{2}(\nabla{\hat {\bf P}})^2+\frac{a_1}{2}({\hat  P}_1^2+{\hat P}_2^2+{\hat P}_3^2)
+\frac{a_{11}}{4}({\hat P}_1^4+{\hat  P}_2^4+{\hat  P}_3^4)\nonumber\\
&+\frac{a_{111}}{6}({\hat P}_1^6+{\hat P}_2^6+{\hat  P}_3^6)
+a_{12}({\hat P}_1^2{\hat P}_2^2+{\hat  P}_2^2{\hat  P}_3^2+{\hat P}_1^2{\hat P}_3^2)
+a_{123}{\hat P}_1^2{\hat P}_2^2{\hat P}_3^2\nonumber\\
&+a_{112}\big[{\hat P}_1^4({\hat P}_2^2+{\hat P}_3^2)+{\hat P}_2^4({\hat P}_1^2+{\hat P}_3^2)+{\hat P}_3^4({\hat P}_1^2+{\hat P}_2^2)\big]\Big]\;d{\bf x}.
\end{align}
Here ${\hat P}_i$ denotes the $i$\textsuperscript{th} component of the polarization field ${\hat {\bf P}}$, $m_p$ is the effective inertia (mass) of the polarization field inherited from the lattice dynamics, $g$ is the stiffness coefficient of the polarization field, the coefficients $a_i$, $a_{ij}$, and $a_{ijk}$ characterize the effective potential of the polarization field. The quadratic coefficient $a_1$ describes the harmonic restoring force of the soft mode and determines the onset of the ferroelectric instability. The higher-order coefficients $a_{11}$ and $a_{12}$ represent quartic anharmonic interactions, while $a_{111}$, $a_{112}$, and $a_{123}$ describe sixth-order anharmonic couplings. These anharmonic terms stabilize the condensed ferroelectric state and determine the magnitude of the spontaneous polarization as well as the nature of the collective modes. 

The ferroelectric phase selects a spontaneous polarization direction. For tetragonal \ce{PbTiO3}, the equilibrium polarization is along the crystallographic c-axis, so that the polarization field can be decomposed as
\begin{equation}
{\hat {\bf P}}({\bf r})={\bf P}+\delta{\bf P}({\bf r}),  
\end{equation}
where ${\bf P}$ is the long-range ferroelectric order parameter and $\delta{\bf P}$
is the fluctuation field $\delta{\bf P}$, which can be further separated into transverse and longitudinal components with respect to the ordered polarization direction, 
\begin{eqnarray}
{\bf P}=P{\bf e}_1 ,\qquad
\delta{\bf P}({\bf r})
=
\delta P_{1}({\bf r}){\bf e}_1
+
\delta P_{2}({\bf r}){\bf e}_2
+
\delta P_{3}({\bf r}){\bf e}_3.
\end{eqnarray}
Here $\delta P_1$ denotes the longitudinal fluctuation parallel to the ordered polarization direction. It changes the magnitude of the ferroelectric order parameter and therefore corresponds to the Higgs, mode of the polarization field. $\delta P_{2}$ and $\delta P_{3}$ describe the transverse fluctuations perpendicular to the spontaneous polarization. These modes correspond to local rotations of the polarization direction and therefore represent pseudo-Goldstone modes, while the tetragonal crystal anisotropy explicitly breaks the continuous rotational symmetry and gives them finite excitation gaps. 

Following the derivation procedure of the self-consistent renormalization~\cite{yang2024self}, one can obtain the energy spectra of the collective polarization modes as
\begin{equation}
\hbar\omega_{\rm A/G}(q)
=
\sqrt{
\Delta_{\rm A/G}^2
+{\hbar^2 g q^2}/{m_p}
}=\sqrt{
\Delta_{\rm A/G}^2
+{\hbar^2 v^2q^2}
},
\end{equation}
where $\omega_{\rm A}(q)$ and $\omega_{\rm G}(q)$ denote the dispersions of the Higgs mode and pseudo-Goldstone mode, respectively. The corresponding excitation gaps $\Delta_{\rm A}$ and $\Delta_{\rm G}$  are derived as 
\begin{eqnarray}
\Delta_A^2&=&{\hbar^2}/{m_p}\big[a_1+3a_{11}P^2+5a_{111}P^4+3a_{11}\langle\delta{P}^2_1\rangle+2a_{12}\big(\langle\delta{P}^2_2\rangle+\langle\delta{P}^2_3\rangle\big)\nonumber\\
&&+a_{111}\big(30P^2\langle\delta{P}^2_1\rangle+5\langle\delta{P}^4_1\rangle\big)+a_{112}\big(2\langle\delta{P}^4_2\rangle+2\langle\delta{P}^4_3\rangle+12P^2\langle\delta{P}^2_2\rangle\nonumber\\
&&+12\langle\delta{P}^2_1\rangle\langle\delta{P}^2_2\rangle+12P^2\langle\delta{P}^2_3\rangle+12\langle\delta{P}^2_1\rangle\langle\delta{P}^2_3\rangle\big)+2a_{123}\langle\delta{P}^2_2\rangle\langle\delta{P}^2_3\rangle\big],\\
\Delta_G^2&=&{\hbar^2}/{m_p}\big[a_1+3a_{11}\langle\delta{P}^2_2\rangle+5a_{111}\langle\delta{P}^4_2\rangle+2a_{12}\big(P^2+\langle\delta{P}^2_1\rangle+\langle\delta{P}^2_3\rangle\big)\nonumber\\
&&+a_{112}\big(2P^4+12P^2\langle\delta{P}^2_1\rangle+2\langle\delta{P}^4_1\rangle+12\langle\delta{P}^2_2\rangle\langle\delta{P}^2_3\rangle+12P^2\langle\delta{P}^2_2\rangle\nonumber\\
&&+12\langle\delta{P}^2_1\rangle\langle\delta{P}^2_2\rangle+2\langle\delta{P}^4_3\rangle\big)+2a_{123}\big(\langle\delta{P}^2_1\rangle\langle\delta{P}^2_3\rangle+P^2\langle\delta{P}^2_3\rangle\big)\big].
\end{eqnarray}
The fluctuation amplitudes are determined self-consistently from the Bose occupation of the collective modes,
\begin{equation}
\langle{\delta}P_1^2\rangle
=
\int\frac{\hbar^2 \text{d}{q}}{(2\pi)^3}
\frac{n_B(\hbar\omega_A)}
     {m_p\omega_A},\quad
\langle{\delta}P_2^2\rangle=\langle{\delta}P_3^2\rangle=
\int\frac{\hbar^2 \text{d}{q}}{(2\pi)^3}
\frac{n_B(\hbar\omega_G)}
     {m_p\omega_G},
\end{equation}
where $n_B(x)$ is the Bose-Einstein distribution function. Under the Gaussian (Hartree) approximation, the fourth-order moments satisfy $\langle \delta P_i^4\rangle=3\langle \delta P_i^2\rangle^2$~\cite{peskin2018introduction}.

The spontaneous polarization is determined by minimizing the free energy of the global polarization $P(T)$ at finite temperatures~\cite{yang2024self},
\begin{equation}
\mathcal{F}(T)=\int d{\bf r}
\left[
\frac{\alpha(T)}{2}P^2
+\frac{\beta(T)}{4}P^4
+\frac{\lambda}{6}P^6
\right],
\label{freeenergy}
\end{equation}
where the free-energy parameters $\alpha(T)$ and $\beta(T)$ are renormalized by the thermal excitation of the Higgs and pseudo-Goldstone modes:
\begin{eqnarray}
\alpha&=&a_1+3a_{11}\langle{\delta}P_1^2\rangle+2a_{12}\big(\langle{\delta}P_2^2\rangle+\langle{\delta}P_3^2\rangle\big)+5a_{111}\langle{\delta}P_1^4\rangle+2a_{123}\langle{\delta}P_2^2\rangle\langle{\delta}P_3^2\rangle\nonumber\\
&&+a_{112}\big(12\langle{\delta}P_1^2\rangle\langle{\delta}P_2^2\rangle+12\langle{\delta}P_1^2\rangle\langle{\delta}P_3^2\rangle+2\langle{\delta}P_2^4\rangle+2\langle{\delta}P_3^4\rangle\big),\\
\beta&=&a_{11}+10a_{111}\langle{\delta}P_1^2\rangle+4a_{112}\big(\langle{\delta}P_2^2\rangle+\langle{\delta}P_3^2\rangle\big),\\
\lambda&=&a_{111}.
\end{eqnarray}
The renormalized free-energy coefficients, the collective-mode gaps, and the fluctuation amplitudes are mutually coupled, forming a closed set of self-consistent equations. Starting from the zero-temperature parameters $a_i$, $a_{ij}$, $a_{ijk}$, $g$, and $m_p$, one can iteratively solve these equations at each temperature. Consequently, all finite-temperature properties, including the spontaneous polarization, collective-mode spectra, dielectric response, and the ferroelectric phase-transition temperature $T_c$, are obtained. 

We applied this self-consistent framework to ferroelectric PbTiO$_3$ and evaluated its finite-temperature properties (\textbf{Supplemental Figure SI-16}). The calculated polarization, dielectric response, lattice strain, and pyroelectric coefficient are all in quantitative agreement with independent experimental measurements. Based on the resulting free-energy landscape and collective fluctuations, we further calculated the collective-mode spectrum without introducing any additional adjustable parameters. The predicted excitation energies are consistent with the Raman and EELS excitations observed in the present work (\textbf{Supplemental Figure SI-16(d)}).\\

\noindent\textbf{Phase-Field Simulations} Phase-field simulations of a lead titanate thin-film were run using Q-POP-FerroDyn, an in-house phase-field solver. The canonical formulation of the phase-field model of ferroelectric materials was used, where the thermodynamics of the material is described using the following free-energy density:
\begin{equation}
    \mathcal{F} = \int_{V} \left(\mathcal{F}^{Landau} + \mathcal{F}^{gradient} + \mathcal{F}^{elastic} + \mathcal{F}^{electric}\right) dV.
\end{equation}
Details about the form of the free-energy can be found in the literature \cite{liEffectSubstrateConstraint2002, chenPhaseFieldMethodPhase2008}. Within the phase-field model, the ferroelectric polarization $P_i$ is evolved using the time-dependent Ginzburg-Landau (TDGL) equation, ${\partial P_{i}}/{\partial t} = -L{\delta \mathcal{F}}/{\delta P_{i}}$, subject to the mechanical (stress, $\sigma_{ij}$) and electrostatic equilibrium (electric potential, $\phi$) conditions: $\nabla\cdot\sigma_{ij}= 0$, and $\nabla\cdot\left(\varepsilon_0\kappa_{ij}\nabla\phi\right) = \nabla\cdot P_{i}$. The system size of the simulations is $256\Delta x\times256\Delta y\times 60\Delta z$, and the cell sizes are $\Delta x = \Delta y = \Delta z = 1 $ nm. The substrate is 20 cells thick and the film is 36 cells thick. The time-step was set to 50 ps, and the simulations were run starting from a randomly-initialized polarization state until the changes in energy and polarization between time-steps decreased to within an acceptable convergence criterion. Each simulation is identical with the exception of the temperature of the system. The domain structures and mean polarization at equilibrium were obtained as a function of temperature to estimate the transition temperatures of the PbTiO$_3$ as grown on a DyScO$_3$ substrate.  Material parameters used for the simulations can be found in {\textbf{Supplemental Table 2}}.\\

\textbf{Thin Film Growth} \SI{100}{\nm} \ce{PbTiO3} thin films were synthesized on titanium dioxide-terminated single-crystal \ce{DyScO3}(110)$_o$ substrates via RHEED-assisted pulsed-laser deposition (krypton-fluoride laser). \ce{PbTiO3} layers were grown at \SI{620}{\degreeCelsius} in \SI{100}{\milli\torr} oxygen pressure. The laser fluence was \SI{1.1}{\joule\per\cm\squared} with a repetition rate of \SI{10}{\hertz}. RHEED was used during deposition to maintain layer-by-layer growth for \ce{PbTiO3}. The specular RHEED spot was used to monitor the RHEED oscillations (\textbf{Supplemental Figure SI 1}). After deposition, thin films were annealed for \SI{10}{\minute} in atmospheric oxygen pressure to promote full oxidation and then cooled down to room temperature at that oxygen pressure. \\

\noindent\textbf{Raman Spectroscopy} Unpolarized and polarized (VV and VH) Raman measurements were performed in a backscattering geometry. A \SI{150}{\milli\watt} Coherent Sapphire SF laser was used with a \SI{532}{\nm} laser wavelength and a 100$\times$ objective lens. The laser power was set at \SI{1.3}{\milli\watt} and was focused to a spot size of approximately \SI{1}{\um}. Data were collected using an Andor 750 spectrometer combined with an iDus Backthinned charge-coupled device detector. In the VV polarization configuration, the incident and scattered light polarizations are parallel, whereas they are perpendicular in the VH configuration. Temperature-dependent Raman measurements were performed using a Linkam HFS350VE-PB4 stage operated at atmospheric pressure in an air environment. Raman spectra were acquired at four different temperatures, ranging from room temperature to \SI{150}{\degree}C, under unpolarized, VV, and VH configurations. At each temperature, data were collected only after the sample temperature had stabilized. All Raman spectra were fitted using pseudo-Voigt functions to determine the peak positions of the corresponding Raman modes.
 \\

\noindent\textbf{TEM Sample Preparation} The cross-sectional TEM sample of thin-film \ce{PbTiO3} grown on \ce{DyScO3} was prepared by a focused ion beam (FIB) technique using a Helios 5UX SEM/FIR with a gallium ion source. At the initial stage of the FIB experiment, a ${\sim}$\SI{100}{\nm} amorphous carbon protective layer was deposited using electron-beam induced deposition, followed by a ${\sim}$\SI{100}{\nm} platinum layer which was deposited under the same conditions. Subsequently, an additional ${\sim}$\SI{1000}{\nm} platinum layer was deposited by ion beam-induced deposition to provide enhanced protection against gallium ion-induced damage during milling. The final polishing step in the FIB preparation was carried out at progressively reduced accelerating voltages, with the ion beam energy gradually lowered to \SI{2}{\kilo\volt}. Furthermore, to clean the gallium ion-induced damage in FIB, a precision ion polishing system II (PIPS-II, Gatan) was employed with an argon ion source operated at an accelerated voltage of 500--\SI{1000}{\electronvolt}. \\

\noindent\textbf{Preliminary Structural Characterization.} Weak-beam dark-field imaging was performed on a Thermo Fisher microscope operated at \SI{200}{\kilo\volt}. After initial bright-field alignment, the g=200 diffracted beam was excited slightly off the exact Bragg condition (by about 0.1–0.2g), and the objective lens excitation was tuned to satisfy the weak-beam condition. Images were acquired with an exposure time of \SI{5}{\second}, using a selected area aperture centered on g=200 to isolate the diffracted spot. Atomic resolution images of \ce{PbTiO3} were taken at \SI{300}{\kilo\volt} in the Thermo Fisher microscope using a \SI{30}{\milli\radian} convergence angle. Tetragonality was determined by gaussian fitting of atoms at each STEM probe position using Python-based implementation in Atomap~\cite{nord2017atomap}. \\

\noindent\textbf{EELS Data Acquisition} STEM-EELS experiments were performed at an accelerating voltage of \SI{60}{\kilo\volt} using a probe corrected NION UltraSTEM, equipped with a monochromator and and a direct electron detector (DECTRIS ELA). Both the convergence semi-angle and collection semi-angle were chosen to be \SI{19} {\milli\radian} and \SI{20} {\milli\radian} respectively, and energy dispersion was \SI{1}{\meV} per channel. Before performing off-axis STEM-EELS, the relative rotation between HAADF-STEM channels and the ronchigram was calibrated. To achieve the off-axis criterion, the central diffraction spot was displaced by approximately \SI{40}{\milli\radian}, moving it completely outside the EELS entrance aperture along four crystallographic directions (conjugate pairs of $\langle001\rangle_\text{pc}$ denoted as 1,2 directions and $\langle100\rangle_\text{pc}$ denoted as 3,4 directions). The direction specific STEM-EEL spectra was obtained by taking an average of 1,2 along  $\langle001\rangle_\text{pc}$  and 3,4 along $\langle100\rangle_\text{pc}$. EELS maps were acquired as a single frame over an $8\times10$~\si{\nm\squared} region with a pixel size of \SI{0.39}{\nm} and a \SI{400}{\ms} time per pixel. The zero-loss peak full width at half maximum or the energy resolution was between \SI{15}{\milli\electronvolt} for all spectra under the off-axis conditions. The data processing is explained in \textbf{Supplemental Note 4}.

Supplemental Information and Methods can be requested from the corresponding authors upon reasonable request.



\bibliography{apssamp}

\end{document}